\documentclass[a4paper, twocolumn]{article}
\usepackage{graphicx}
\usepackage{amsmath}

\begin{document}

\title{The coherent information of Pauli channels with coded inputs}
\author{Xiao-yu Chen Li-zhen Jiang\\
{\small {College of Information and Electronic Engineering, Zhejiang
Gongshang University,Hangzhou,310018,China }}}
\date{}
\maketitle

\begin{abstract}
The calculating of the coherent information is a fundamental step in
obtaining the quantum capacity of a quantum channel. We introduce orthogonal
and complete code basis to evaluate the coherent information per channel use
when the input is the maximal mixture of stabilizer codewords. In the code
basis, the output density matrix is diagonal, the joint output of the system
and the auxiliary is block diagonal. The coherent information is worked out
by counting the weights of error operators.

\end{abstract}


\section{Introduction}

The basic issue in quantum information theory is quantum coding theorem.
After ten year's efforts, quantum coding theorem had at last been proven.
The rate of faithfully transmitting quantum information per use of quantum
channel is limited by quantum capacity, the capacity is asymptotically
achievable \cite{Lloyd} \cite{Barnum} \cite{Horodecki} \cite{Devetak}.
Quantum capacity is the maximization of coherent information \cite
{Schumacher} over all input states. Unfortunately, since coherent
information is non-additive \cite{DiVincenzo}, quantum capacity in single
letter form is not available except for degradable \cite{Devetak0} or
anti-degradable channels. Regulation is need, that is, block input with
infinitive number of qubits should be used to calculate quantum capacity in
general. We may only obtain the lower bound of quantum capacity. The
obstacle of obtaining the coherent information with other multipartite input
state is obvious, the dimension of the state increases exponentially with
the the number of the input qubits, making the calculation of the output
entropy and the entropy exchange (thus the coherent information) an awful
work. We will greatly reduce the complexity of diagonalizing the output
density matrix by introducing quantum error-correcting code (QECC) as the
input state.

\section{QECC and Pauli Channels}

The theory of QECCs was established more than a decade ago as the tool for
fighting decoherence in quantum computers and quantum communication systems
\cite{Shor}. Maybe the most impressive development in quantum
error-correction theory is the use of the stabilizer formalism\cite
{Calderbank} \cite{Gottesman} \cite{Gottesman0} \cite{Nielsen}. The power of
the stabilizer formalism comes from the clever use of group theory. The $n$%
-fold Pauli operators $\{I,X,Y,Z\}^{\otimes n}$ together with the possible
overall factors $\pm 1,\pm i$ form a group $G_n$ under multiplication, the $%
n $-fold Pauli group. Suppose $S$ is an abelian subgroup of $G_n$.
Stabilizer coding space $T$ is the simultaneous $+1$ eigenspace of all
elements of $S$, $T=\{\left| \psi \right\rangle :$ $M\left| \psi
\right\rangle =\left| \psi \right\rangle ,\forall M\in S\}.$ For an $%
[[n,k,d]]$ stabilizer code, which encodes $k$ logical qubits into $n$
physical qubits, $T$ has dimension $2^k$ and $S$ has $2^{n-k\text{ }}$%
elements. The generators of $S$ are denoted as $M_i$ $(i=1,\ldots ,n-k)$
which are Hermitian. There are many elements in $G_n $ that commute with
every elements of $S$ but not actually in $S.$ The set of elements in $G_n$
that commute with all of $S$ is defined as the centralizer $C(S)$ of $S$ in $%
G_n$. Clearly $S\subset C(S).$

Denote $\Omega =\prod_i(I+M_i).$ Due to the properties that the elements of
stabilizer group $S$ commute and $M_i^2=I,$ it follows $M_j\prod_i(I+M_i)=%
\prod_i(I+M_i).$ Further, we have $\Omega M_j\Omega =\Omega ^2=2^{n-k}\Omega
.$ For error operator $E_a$ that anti-commutes with at least one of the
generators $M_i,$ we have
\begin{equation}
\Omega E_a\Omega =0,  \label{wee1}
\end{equation}
this is due to $(I+M_i)(I-M_i)=I-M_i^2=0,$ the operator factor $(I-M_i)$
comes from $E_a(I+M_i)=(I-M_i)E_a$ if $E_a$ anti-commute with $M_i$.

In Krauss representation, Pauli channel map $\mathcal{E}$ acting on qubit
state $\rho $ can be written as $\mathcal{E}\left( \rho \right) =f\rho
+p_xX\rho X+p_yY\rho Y+p_zZ\rho Z,$ where $p_{x(y,z)}\in [0,1]$ are the
probabilities, $f=1-p_x-p_y-p_z$ $\in [0,1]$ is the fidelity of the channel.
For depolarizing channel, $p_x=p_y=p_z=p,$ $f=1-3p.$ The total error
probability is $3p.$  For $n$ use of depolarizing channels with $n$ qubits
input state $\rho $, we have the output state $\rho ^{\prime }=\mathcal{E}%
^{\otimes n}\left( \rho \right) =\sum_a\eta _aE_a\rho E_a^{\dagger },$ with $%
\eta _a=f^{n-i-j-l}p_x^ip_y^jp_z^l$ for $E_a=X^iY^jZ^l.$ The purification of
$\rho $ is $\left| \Psi \right\rangle ,$ we have the joint output state $%
\rho _e=(\mathcal{E}^{\otimes n}\otimes I^{\otimes n})\left( \left| \Psi
\right\rangle \left\langle \Psi \right| \right) ,$ whose entropy is the
entropy exchange. The coherent information is $I_c=S(\rho ^{\prime })-S(\rho
_e),$ where $S(\cdot )$ is the von Neumann entropy.

\section{The coherent information of depolarizing channel with coded input}

\subsection{The [[5,1,3]] code}

To illustrate non-zero capacity even for zero fidelity of Pauli channel, we
first consider the example of $[[5,1,3]]$ QECC as the input state to the
depolarizing channel. The stabilizer code has four generators $%
M_1=X_1Z_2Z_3X_4,$ $M_2=X_2Z_3Z_4X_5,$ $M_3=X_1X_3Z_4Z_5,$ $M_4=Z_1X_2X_4Z_5$%
. The codewords are $\left| \overline{0}\right\rangle =\frac 14\Omega \left|
00000\right\rangle ,\left| \overline{1}\right\rangle =\frac 14\overline{X}%
\Omega \left| 00000\right\rangle ,$ where $\overline{X}=XXXXX$, $\overline{X}%
\subset C(S)\backslash S.$ Another useful operator in $C(S)\backslash S$ is $%
\overline{Z}=ZZZZZ,$ which anti-commutes with $\overline{X}.$ The input
state $\rho $ is chosen to be $\frac 12(\left| \overline{0}\right\rangle
\left\langle \overline{0}\right| +\left| \overline{1}\right\rangle
\left\langle \overline{1}\right| )$ with maximal input entropy in the
logical qubit basis $\left| \overline{0}\right\rangle $ and $\left|
\overline{1}\right\rangle $. To evaluate the eigenvalues of output state and
joint output state of the system and the auxiliary, we introduce the code
basis: $\left| J\right\rangle =\Lambda _J\left| \overline{0}\right\rangle $ $%
(J=0,\ldots ,31),$ where $\Lambda _J=I,X_m$ $(m=1,\ldots ,5),Y_m,Z_m$ ,$%
\overline{X},X_m\overline{X}$,$Y_m\overline{X},Z_m\overline{X}.$ $.$ Note
that $[[5,1,3]]$ can correct any single qubit error, the basis are
orthonormal. For input code state $\left| \overline{0}\right\rangle ,$ the
elements of output density matrix are $\left\langle J\right| \sum_a\eta
_aE_a\left| \overline{0}\right\rangle \left\langle \overline{0}\right|
E_a^{\dagger }\left| K\right\rangle .$ Denote $\left| \mathbf{0}%
\right\rangle =\left| 00000\right\rangle ,$ the factor $\left\langle
J\right| E_a\left| \overline{0}\right\rangle =\frac 1{16}\left\langle
\mathbf{0}\right| \Omega \Lambda _J^{\dagger }E_a\Omega \left| \mathbf{0}%
\right\rangle $ is $0$ when $\Lambda _J^{\dagger }E_a\notin S+S\overline{Z}$
(up to $\pm 1$,$\pm i$ factors, which have no effect in channel mapping and
are omitted hereafter) $,$ and is $1$ when $\Lambda _J^{\dagger }E_a\in S+S%
\overline{Z}$ , according to the following reasons. We have three cases: (i)
$\Lambda _J^{\dagger }E_a\notin C(S),$ then $\Lambda _J^{\dagger }E_a$
anti-commutes with some generators of $S$, by Eq. (\ref{wee1}), so $%
\left\langle J\right| E_a\left| \overline{0}\right\rangle =0$; (ii) $\Lambda
_J^{\dagger }E_a\in C(S)/(S+S\overline{Z})$, equivalently $\Lambda
_J^{\dagger }E_a\in \overline{X}(S+S\overline{Z}).$ Then there is some $M\in
S\ $such that $\frac 1{16}\left\langle \mathbf{0}\right| \Omega \Lambda
_J^{\dagger }E_a\Omega \left| \mathbf{0}\right\rangle =$ $\frac
1{16}\left\langle \mathbf{0}\right| \Omega M\Omega \overline{X}\left|
\mathbf{0}\right\rangle =$ $\frac 1{16}\left\langle \mathbf{0}\right| \Omega
\overline{X}\Omega \left| \mathbf{0}\right\rangle =$ $\left\langle \overline{%
0}\right| \left. \overline{1}\right\rangle =0$; (iii) $\Lambda _J^{\dagger
}E_a\in S+S\overline{Z}$, then there is some $M\in S\ $such that $%
\left\langle J\right| E_a\left| \overline{0}\right\rangle =$ $\frac
1{16}\left\langle \mathbf{0}\right| \Omega M\Omega \left| \mathbf{0}%
\right\rangle =$ $\frac 1{16}\left\langle \mathbf{0}\right| \Omega \Omega
\left| \mathbf{0}\right\rangle =1$. If $\Lambda _J^{\dagger }E_a\in S+S%
\overline{Z}$ and $E_a^{\dagger }\Lambda _K\in S+S\overline{Z},$ we have
their product $\Lambda _J^{\dagger }E_aE_a^{\dagger }\Lambda _K=\Lambda
_J^{\dagger }\Lambda _K\in S+S\overline{Z}$ , which is only possible when $%
J=K$, since our $\Lambda _J$ is so chosen that each of which is the head of
one of cosets $G_5/(S+S\overline{Z}).$ Thus the state $\mathcal{E}^{\otimes
5}(\left| \overline{0}\right\rangle \left\langle \overline{0}\right| )$ is
diagonal in the code basis. Similar result can be found for $\mathcal{E}%
^{\otimes 5}(\left| \overline{1}\right\rangle \left\langle \overline{1}%
\right| ).$ The output state in the representation of the code basis is
diagonalized.
\begin{equation}
\rho _{JK}^{\prime }=\frac 12\delta _{JK}(\sum_a\eta _a+\sum_{a^{\prime
}}\eta _a^{\prime }),\text{ }  \label{wee2}
\end{equation}
with the conditions of $\Lambda _J^{\dagger }E_a$ $\in S+S\overline{Z}$ for
the first term, and $\overline{X}\Lambda _J^{\dagger }E_a^{\prime }\in S+S%
\overline{Z}$ for the second term at the right hand side $.$

The purification of the input state $\rho $ could be $\left| \Psi
\right\rangle =\frac 1{\sqrt{2}}(\left| \overline{0}\right\rangle \left|
\overline{0}_A\right\rangle +\left| \overline{1}\right\rangle \left|
\overline{1}_A\right\rangle ),$ where the first logical qubit is for the
system , the second logical qubit is for the auxiliary and denoted by the
subscript $a$. The joint output state of the system and the auxiliary is $%
\rho _e=(\mathcal{E}^{\otimes 5}\otimes I^{\otimes 5})\left( \left| \Psi
\right\rangle \left\langle \Psi \right| \right) $. An obvious basis for the
joint output density matrix is $\left| J\right\rangle \left| \overline{0}%
_A\right\rangle ,\left| J\right\rangle \left| \overline{1}_A\right\rangle .$
The non-zero elements are $\left\langle J\right| \left\langle \overline{0}%
_A\right| \rho _e\left| J\right\rangle \left| \overline{0}_A\right\rangle ,$
$\left\langle J\right| \left\langle \overline{0}_A\right| \rho _e\left|
J^{\prime }\right\rangle \left| \overline{1}_A\right\rangle ,$ $\left\langle
J^{\prime }\right| \left\langle \overline{1}_A\right| \rho _e\left|
J\right\rangle \left| \overline{0}_A\right\rangle ,\left\langle J^{\prime
}\right| \left\langle \overline{1}_A\right| \rho _e\left| J^{\prime
}\right\rangle \left| \overline{1}_A\right\rangle $ $,$ where $J^{\prime
}=mod(J+16,32).$ The reason can be seen from the calculating $\left\langle
J\right| \left\langle \overline{0}_A\right| \rho _e\left| K\right\rangle
\left| \overline{1}_A\right\rangle $ for example. We have $\left\langle
J\right| \left\langle \overline{0}_A\right| \rho _e\left| K\right\rangle
\left| \overline{1}_A\right\rangle =$ $\frac 12\left\langle J\right|
\mathcal{E}^{\otimes 5}(\left| \overline{0}\right\rangle \left\langle
\overline{1}\right| )\left| K\right\rangle $ $=\frac 12\sum_a\eta
_a\left\langle J\right| E_a\left| \overline{0}\right\rangle \left\langle
\overline{1}\right| E_a^{\dagger }\left| K\right\rangle ,$which is nonzero
when $\Lambda _J^{\dagger }E_a\in S+S\overline{Z}$ and $E_a^{\dagger
}\Lambda _K\overline{X}\in S+S\overline{Z},$ thus $\Lambda _J^{\dagger
}E_aE_a^{\dagger }\Lambda _K\overline{X}=\Lambda _J^{\dagger }\Lambda _K%
\overline{X}\in S+S\overline{Z},$ hence $K=mod(J+16,32)=J^{\prime }.$ In the
basis of $\left| J\right\rangle \left| \overline{0}_a\right\rangle ,\left|
J\right\rangle \left| \overline{1}_a\right\rangle ,$ by rearranging the
subscripts, the matrix $\rho _e$ can be decomposed to the direct summation
of $32$ submatrices. $\rho _e=\oplus _{J=0}^{31}\rho _{eJ}$, with
\begin{equation}
\rho _{eJ}=\left[
\begin{array}{ll}
\left\langle J\right| \left\langle \overline{0}_A\right| \rho _e\left|
J\right\rangle \left| \overline{0}_A\right\rangle & \left\langle J\right|
\left\langle \overline{0}_A\right| \rho _e\left| J^{\prime }\right\rangle
\left| \overline{1}_A\right\rangle \\
\left\langle J^{\prime }\right| \left\langle \overline{1}_A\right| \rho
_e\left| J\right\rangle \left| \overline{0}_A\right\rangle & \left\langle
J^{\prime }\right| \left\langle \overline{1}_A\right| \rho _e\left|
J^{\prime }\right\rangle \left| \overline{1}_A\right\rangle
\end{array}
\right] .
\end{equation}
We have $\left\langle J\right| \left\langle \overline{0}_A\right| \rho
_e\left| J\right\rangle \left| \overline{0}_A\right\rangle =$ $\frac
12\left\langle J\right| \mathcal{E}^{\otimes 5}(\left| \overline{0}%
\right\rangle \left\langle \overline{0}\right| )\left| J\right\rangle =$ $%
\frac 12\sum_a\eta _a\left\langle J\right| E_a\left| \overline{0}%
\right\rangle \left\langle \overline{0}\right| E_a^{\dagger }\left|
J\right\rangle =$ $\frac 12\sum_{E_a\in \Lambda _J(S+S\overline{Z})}\eta _a,$
and $\left\langle J^{\prime }\right| \left\langle \overline{1}_A\right| \rho
_e\left| J^{\prime }\right\rangle \left| \overline{1}_A\right\rangle
=\left\langle J\right| \left\langle \overline{0}_A\right| \rho _e\left|
J\right\rangle \left| \overline{0}_A\right\rangle ;$ $\left\langle J\right|
\left\langle \overline{0}_A\right| \rho _e\left| J^{\prime }\right\rangle
\left| \overline{1}_A\right\rangle =\frac 12\sum_a\eta _a\left\langle
J\right| E_a\left| \overline{0}\right\rangle \left\langle \overline{1}%
\right| E_a^{\dagger }\left| J^{\prime }\right\rangle .$ Notice that $%
\left\langle \overline{1}\right| E_a^{\dagger }\left| J^{\prime
}\right\rangle =\left\langle \overline{0}\right| \overline{X}E_a^{\dagger
}\Lambda _{J^{\prime }}\left| \overline{0}\right\rangle =$ $\left\langle
\overline{0}\right| \overline{X}E_a^{\dagger }\Lambda _J\overline{X}\left|
\overline{0}\right\rangle ,$ if $\Lambda _J^{\dagger }E_a\in S,$ we have $%
\overline{X}E_a^{\dagger }\Lambda _J\overline{X}=E_a^{\dagger }\Lambda _J,$
while for $\Lambda _J^{\dagger }E_a\in S\overline{Z},$ we have $\overline{X}%
E_a^{\dagger }\Lambda _J\overline{X}=-E_a^{\dagger }\Lambda _J.$ Thus
\begin{equation}
\rho _{eJ}=\frac 12\left[
\begin{array}{ll}
\eta _{eJ}+\eta _{oJ} & \eta _{eJ}-\eta _{oJ} \\
\eta _{eJ}-\eta _{oJ} & \eta _{eJ}+\eta _{oJ}
\end{array}
\right] ,
\end{equation}
with
\begin{eqnarray}
\eta _{eJ} &=&\sum_{E_a\in \Lambda _JS}\eta _a,  \label{wee3} \\
\eta _{oJ} &=&\sum_{E_a\in \Lambda _JS\overline{Z}}\eta _a.  \label{wee4}
\end{eqnarray}
Thus the eigenvalues of $\rho _{eJ}$ are $\eta _{eJ}$ and $\eta _{oJ}.$ We
can obtain from (\ref{wee2}) that the eigenvalues of $\rho ^{\prime }$ are
\begin{equation}
\lambda _J=\frac 12(\eta _{oJ}+\eta _{eJ}+\eta _{oJ^{\prime }}+\eta
_{eJ^{\prime }}).  \label{wee5}
\end{equation}
The average coherent information per channel use thus is
\begin{equation}
\overline{I_c}=\frac 15\sum_{J=0}^{32}(-\lambda _J\log _2\lambda _J+\eta
_{oJ}\log _2\eta _{oJ}+\eta _{eJ}\log _2\eta _{eJ}).  \label{wee6}
\end{equation}

The eigenvalues $\eta _{eJ}$ ,$\eta _{oJ}$ can be obtained by counting the
weights of all the operators of $\Lambda _JS$ and $\Lambda _JS\overline{Z},$
respectively. The weight of an error operator is the number of qubits on
which it differs from the identity. An error $E_a\in \Lambda _JS$ with
weight $j$ will contribute $\eta _a=f^{n-j}p^j$ to the eigenvalue of $\eta
_{eJ},$ similarly, An error $E_a\in \Lambda _JS\overline{Z}$ with weight $j$
will contribute $\eta _a=f^{n-j}p^j$ to the eigenvalue of $\eta _{oJ}$. Thus
$\eta _{eJ}$ and $\eta _{oJ}$can be written as $%
\sum_{j=0}^5c_j^{eJ}f^{n-j}p^j$ and $\sum_{j=0}^5c_j^{oJ}f^{n-j}p^j$
,respectively. Furthermore, they are characterized by vector $\mathbf{c}%
_J^e=(c_5^{eJ},c_4^{eJ},c_3^{eJ},c_2^{eJ},c_1^{eJ},c_0^{eJ})$ and $\mathbf{c}%
_J^o=(c_5^{oJ},c_4^{oJ},c_3^{oJ},c_2^{oJ},c_1^{oJ},c_0^{oJ}).$ A detail
counting shows $\mathbf{c}_0^e=(0,15,0,0,0,1),\mathbf{c}_0^o$ $=\mathbf{c}%
_{16}^e=\mathbf{c}_{16}^o=(6,0,10,0,0,0)$. The eigenvalues of the subscripts
$J=1,2,3,4,5,$ $22,23,24,25,26,$ $11,12,13,14,15$ are degenerated and
represented by $\mathbf{c}_1^e=(3,8,4,0,1,0)$ and $\mathbf{c}%
_1^o=(4,6,4,2,0,0);$ also the eigenvalues are degenerate for $%
J=17,18,19,20,21,$ $6,7,8,9,10,$ $27,28,29,30,31$ and represented by $%
\mathbf{c}_{17}^o=\mathbf{c}_{17}^e=\mathbf{c}_1^e.$ The average coherent
information per channel use then is
\begin{eqnarray}
\overline{I_c} &=&\frac 15[-\frac{(\eta _{e0}+3\eta _{o0})}2\log _2\frac{%
(\eta _{e0}+3\eta _{o0})}2  \nonumber \\
&&+\eta _{e0}\log _2\eta _{e0}+3\eta _{o0}\log _2\eta _{o0}]  \nonumber \\
&&+3[-\frac{(\eta _{e1}+3\eta _{o1})}2\log _2\frac{(\eta _{e1}+3\eta _{o1})}2
\nonumber \\
&&+\eta _{e1}\log _2\eta _{e1}+3\eta _{o1}\log _2\eta _{o1}].
\end{eqnarray}

\subsection{The [[7,1,3]] code}

The six stabilizer generators of [[7,1,3]] code are $M_1=X_1X_2X_3X_4,$ $%
M_2=X_1X_2X_5X_6,$ $M_3=X_1X_3X_5X_7,$ $M_4=Z_1Z_2Z_3Z_4,$ $%
M_5=Z_1Z_2Z_5Z_6, $ $M_6=Z_1Z_3Z_5Z_7.$ The bit flip and phase flip
operators for the encoded (logical) qubit are $\overline{X}=X_5X_6X_7$ and $%
\overline{Z}=Z_5Z_6Z_7,$ respectively. Since the code can correct any single
qubit errors of $X,Y,Z$ types and meanwhile it can correct errors $X_iZ_j$ ($%
i\neq j$) type, it is convenient to choose the $128$ coset heads as $\Lambda
_J\ (J=0,\ldots ,127)=I,$ $X_k,(k=1,\ldots ,7),Y_k,Z_k,$ $X_iZ_j$ $(i\neq j;$
$i,j=$ $1,\ldots ,7),$ $\overline{X},$ $X_k\overline{X},(k=1,\ldots ,7),$ $%
Y_k\overline{X},$ $Z_k\overline{X},$ $X_iZ_j\overline{X}$ $(i\neq j;$ $i,j=$
$1,\ldots ,7).$ With $\eta _{eJ}$ and $\eta _{oJ}$ being written as $%
\sum_{j=0}^7c_j^{eJ}f^{n-j}p^j$ and $\sum_{j=0}^7c_j^{oJ}f^{n-j}p^j,$ the
non-zero eigenvalues of the joint output state $\rho _e$ now can be
characterized by vectors $\mathbf{c}_J^e$ and $\mathbf{c}_J^o.$ They are (i)
non-degenerate $\mathbf{c}_0^e=(0,42,0,21,0,0,0,1),$ $\mathbf{c}_0^o=\mathbf{%
c}_{64}^e=\mathbf{c}_{64}^o=(15,0,42,0,7,0,0,0);$ (ii) $7$ fold degenerate
represented by $\mathbf{c}_1^e=(6,24,21,8,4,0,1,0),$ $\mathbf{c}_1^o=$ $%
\mathbf{c}_{65}^o$ $=$ $(11,16,18,16,3,0,0,0),$ $\mathbf{c}%
_{65}^e=(8,19,24,10,0,3,0,0);$(iii) $7$ fold degenerate represented by $%
\mathbf{c}_8^e=\mathbf{c}_1^e,$ $\mathbf{c}_8^o=$ $\mathbf{c}_{65}^e$ $,$ $%
\mathbf{c}_{72}^e=\mathbf{c}_{72}^e=\mathbf{c}_1^o;$(iv) $7$ fold degenerate
represented by $\mathbf{c}_{15}^e=\mathbf{c}_1^e,$ $\mathbf{c}_{15}^o=%
\mathbf{c}_{79}^e$ $=$ $\mathbf{c}_1^o$ $,$ $\mathbf{c}_{79}^o=\mathbf{c}%
_{65}^e;$ (v) $42$ fold degenerate represented by $\mathbf{c}_{23}^e=\mathbf{%
c}_{23}^o=\mathbf{c}_{87}^e=(8,21,20,10,4,1,0,0),$ $\mathbf{c}%
_{87}^o=(9,20,18,12,5,0,0,0).$ The eigenvalues of the output state $\rho
^{\prime }$ are $\lambda _J=\frac 12(\eta _{oJ}+\eta _{eJ}+\eta _{oJ^{\prime
}}+\eta _{eJ^{\prime }}),$ with $J^{\prime }=mod(J+64,128).$ The average
coherent information per channel use then is

\begin{eqnarray}
\overline{I_c} &=&\frac 17[-\frac{(\eta _{e0}+3\eta _{o0})}2\log _2\frac{%
(\eta _{e0}+3\eta _{o0})}2  \nonumber \\
&&+\eta _{e0}\log _2\eta _{e0}+3\eta _{o0}\log _2\eta _{o0}]  \nonumber \\
&&+3[-\frac{(\eta _{e1}+2\eta _{o1}+\eta _{e65})}2\log _2\frac{(\eta
_{e1}+2\eta _{o1}+\eta _{e65})}2  \nonumber \\
&&+\eta _{e1}\log _2\eta _{e1}+2\eta _{o1}\log _2\eta _{o1}+\eta _{e65}\log
_2\eta _{e65}]  \nonumber \\
&&+6[-\frac{(\eta _{e87}+3\eta _{o87})}2\log _2\frac{(\eta _{e87}+3\eta
_{o87})}2  \nonumber \\
&&+\eta _{e87}\log _2\eta _{e87}+3\eta _{o87}\log _2\eta _{o87}].
\end{eqnarray}

\subsection{ The [[8,3,3]] code}

The improvement to the lower bound of quantum capacity comes from $[[8,3,3]]$
code. The code encodes $3$ logical qubits into $8$ physical qubits and
corrects $1$ error. Thus the coding rate is higher than $[[5,1,3]]$ code for
noiseless channel. And it indeed provides a tighter lower bound for the
quantum capacity at high fidelity domain. $[[8,3,3]]$ code has generators $%
M_i$ $(i=1,\ldots ,5)$ of its stabilizer group $S$ , with $%
M_1=\prod_{i=1}^8X_i,$ $M_2=\prod_{i=1}^8Z_i,M_3=X_2X_4Y_5Z_6Y_7Z_8,$ $%
M_4=X_2Z_3Y_4X_6Z_7Y_8,$ $M_5=Y_2X_3Z_4X_5Z_6Y_8$. In the centralizer $C(S)$
there are bit flip and phase flip operators $\overline{X}_j,$ $\overline{Z}%
_j $ $(j=1,2,3)$ for the encoded (logical) qubit, with $\overline{X}%
_1=X_1X_2Z_6Z_8,$ $\overline{X}_2=X_1X_3Z_4Z_7,$ $\overline{X}%
_3=X_1Z_4X_5Z_6,$ $\overline{Z}_1=Z_2Z_4Z_6Z_8,$ $\overline{Z}%
_2=Z_3Z_4Z_7Z_8,$ $\overline{Z}_3=Z_5Z_6Z_7Z_8$. The codewords are $\left|
\overline{k_1k_2k_3}\right\rangle =$ $\frac 1{4\sqrt{2}}\overline{X}_1^{k_1}%
\overline{X}_2^{k_2}\overline{X}_3^{k_3}\Omega \left| \mathbf{0}%
\right\rangle ,$ with $k_i=0,1.$ The channel input state $\rho $ could be
chosen as the equal probability mixture of codeword states, each codeword
has a probability of $\frac 18.$ The basic set of the heads of cosets $%
G_8/(S\times \overline{Z})$ ($\overline{Z}$ is the group with generators $%
\overline{Z}_j$) can be chosen as the correctable single qubit errors $%
X_m,Y_m,Z_m$ and some other two qubit errors such as $X_1X_{i+1}$ ($%
i=1,\ldots ,7$) and the identity $I.$ The number of the elements in the
basic set is $32.$ The whole coset head set is obtained by the
multiplication (at right) of the basic set with group $\overline{X}$ whose
generators are $\overline{X}_j.$ Denote the elements of the coset head as $%
\Lambda _J$ $(J=0,\ldots ,255).$ Suppose the input state is $\rho =\frac
18\sum_{k_1,k_2,k_3=0}^1\left| \overline{k_1k_2k_3}\right\rangle
\left\langle \overline{k_1k_2k_3}\right| $, then the purification of $\rho $
is $\left| \Psi \right\rangle =\frac 1{\sqrt{8}}\sum_{k_1,k_2,k_3=0}^1\left|
\overline{k_1k_2k_3}\right\rangle \left| (\overline{k_1k_2k_3}%
)_A\right\rangle ,$ the joint output of the system and auxiliary is $\rho
_e=(\mathcal{E}^{\otimes 8}\otimes I^{\otimes 8})\left( \left| \Psi
\right\rangle \left\langle \Psi \right| \right) .$ In the basis of $\left|
J\right\rangle =\Lambda _J\left| \overline{000}\right\rangle ,$ we have
\begin{eqnarray}
&&\left\langle K\right| \left\langle \left( \overline{k_1k_2k_3}\right)
_A\right| \rho _e\left| J\right\rangle \left| (\overline{j_1j_2j_3}%
)_A\right\rangle  \nonumber \\
&=&\sum_a\eta _a\left\langle K\right| E_a\left| \overline{k_1k_2k_3}%
\right\rangle \left\langle \overline{j_1j_2j_3}\right| E_a^{\dagger }\left|
J\right\rangle ,
\end{eqnarray}
which is nonzero when $\overline{X}_1^{k_1}\overline{X}_2^{k_2}\overline{X}%
_3^{k_3}\Lambda _K^{\dagger }E_a\in S\times \overline{Z}$ and $E_a^{\dagger
}\Lambda _J\overline{X}_1^{j_1}\overline{X}_2^{j_2}\overline{X}_3^{j_3}\in
S\times \overline{Z},$ thus $\overline{X}_1^{k_1}\overline{X}_2^{k_2}%
\overline{X}_3^{k_3}\Lambda _K^{\dagger }E_aE_a^{\dagger }\Lambda _J%
\overline{X}_1^{j_1}\overline{X}_2^{j_2}\overline{X}_3^{j_3}=$ $\overline{X}%
_1^{k_1}\overline{X}_2^{k_2}\overline{X}_3^{k_3}\Lambda _K^{\dagger }\Lambda
_J\overline{X}_1^{j_1}\overline{X}_2^{j_2}\overline{X}_3^{j_3}$ $\in S\times
\overline{Z}.$ However, $\Lambda _K$ and $\Lambda _J$ are coset heads, so we
have $\Lambda _K=\Lambda _J\overline{X}_1^{k_1+j_1}\overline{X}_2^{k_2+j_2}%
\overline{X}_3^{k_3+j_3}.$ Let $\overline{X}_1^{k_1}\overline{X}_2^{k_2}%
\overline{X}_3^{k_3}\Lambda _K^{\dagger }E_a=S_b\overline{Z}_1^{c_1}%
\overline{Z}_2^{c_2}\overline{Z}_3^{c_3},$ where $S_b$ ($b=0,\ldots ,31$) is
the element of stabilizer group $S$ and $c_i=0,1.$ Then $E_a=\Lambda _K%
\overline{X}_1^{k_1}\overline{X}_2^{k_2}\overline{X}_3^{k_3}S_b\overline{Z}%
_1^{c_1}\overline{Z}_2^{c_2}\overline{Z}_3^{c_3}=$ $\Lambda _J\overline{X}%
_1^{j_1}\overline{X}_2^{j_2}\overline{X}_3^{j_3}S_b\overline{Z}_1^{c_1}%
\overline{Z}_2^{c_2}\overline{Z}_3^{c_3}.$ We have $\left\langle K\right|
E_a\left| \overline{k_1k_2k_3}\right\rangle =$ $\left\langle \overline{000}%
\right| \Lambda _K^{\dagger }\Lambda _K\overline{X}_1^{k_1}\overline{X}%
_2^{k_2}\overline{X}_3^{k_3}S_b\overline{Z}_1^{c_1}\overline{Z}_2^{c_2}%
\overline{Z}_3^{c_3}\overline{X}_1^{k_1}\overline{X}_2^{k_2}\overline{X}%
_3^{k_3}\left| \overline{000}\right\rangle =$ $(-1)^{k_1c_1+k_2c_2+k_3c_3}$
and $\left\langle \overline{j_1j_2j_3}\right| E_a^{\dagger }\left|
J\right\rangle =\left\langle \overline{000}\right| \overline{X}_1^{j_1}%
\overline{X}_2^{j_2}\overline{X}_3^{j_3}\overline{Z}_3^{c_3}\overline{Z}%
_2^{c_2}\overline{Z}_1^{c_1}S_b^{\dagger }\overline{X}_3^{j_1}\overline{X}%
_2^{j_2}\overline{X}_1^{j_3}\Lambda _J^{\dagger }\Lambda _J\left| \overline{%
000}\right\rangle =(-1)^{j_1c_1+j_2c_2+j_3c_3}.$ So that
\begin{eqnarray}
&&\sum_a\eta _a\left\langle K\right| E_a\left| \overline{k_1k_2k_3}%
\right\rangle \left\langle \overline{j_1j_2j_3}\right| E_a^{\dagger }\left|
J\right\rangle  \nonumber \\
&=&\sum_{a^{\prime }}(-1)^{\sum_{i=1}^3(k_i+j_i)c_i}\eta _{a^{\prime }}.
\end{eqnarray}
Where $E_{a^{\prime }}\in \Lambda _J\overline{X}_1^{j_1}\overline{X}_2^{j_2}%
\overline{X}_3^{j_3}S\overline{Z}_1^{c_1}\overline{Z}_2^{c_2}\overline{Z}%
_3^{c_3},$ and $mod(K+32(4k_3+2k_2+k_1),256)=mod(J+32(4j_3+2j_2+j_1),256).$
By rearranging the basis, the joint output state $\rho _e$ can be written in
a block diagonalized form with each block being a $8\times 8$ submatrix. A
detail analysis shows that each $8\times 8$ submatrix can be diagonalized
with Hadamard transformation. The eigenvalues of $\rho _e$ are $\sum_{a\in
\Lambda _JS}\eta _a,$ $\sum_{a\in \Lambda _JS\overline{Z}_1}\eta _a,$ $%
\sum_{a\in \Lambda _JS\overline{Z}_2}\eta _a,$ $\sum_{a\in \Lambda _JS%
\overline{Z}_1\overline{Z}_2}\eta _a,$ $\sum_{a\in \Lambda _JS\overline{Z}%
_3}\eta _a,$ $\sum_{a\in \Lambda _JS\overline{Z}_1\overline{Z}_3}\eta _a,$ $%
\sum_{a\in \Lambda _JS\overline{Z}_2\overline{Z}_3}\eta _a,$ $\sum_{a\in
\Lambda _JS\overline{Z}_1\overline{Z}_2\overline{Z}_3}\eta _a.$ The total
number of the nonzero eigenvalues of $\rho _e$ is $256\times 8=2048.$ A
detail counting shows that each of the eigenvalue should be on of the $\xi
_i $ $(i=1,\ldots ,14),$ where $\xi _i=\sum_{j=0}^8c_{i,j+1}f^{n-j}p^j$ with
\[
c=\left[
\begin{array}{lllllllll}
1 & 0 & 0 & 0 & 0 & 0 & 28 & 0 & 3 \\
0 & 0 & 0 & 0 & 6 & 0 & 20 & 0 & 6 \\
0 & 0 & 0 & 2 & 0 & 12 & 0 & 18 & 0 \\
0 & 0 & 1 & 0 & 1 & 8 & 11 & 8 & 3 \\
0 & 0 & 0 & 0 & 4 & 8 & 8 & 8 & 4 \\
0 & 0 & 0 & 2 & 2 & 4 & 12 & 10 & 2 \\
0 & 0 & 0 & 1 & 3 & 6 & 10 & 9 & 3 \\
0 & 1 & 0 & 0 & 0 & 7 & 14 & 8 & 2 \\
0 & 0 & 0 & 1 & 2 & 8 & 10 & 7 & 4 \\
0 & 0 & 0 & 2 & 1 & 6 & 12 & 8 & 3 \\
0 & 0 & 1 & 0 & 2 & 6 & 11 & 10 & 2 \\
0 & 0 & 0 & 1 & 4 & 4 & 10 & 11 & 2 \\
0 & 0 & 0 & 0 & 3 & 10 & 8 & 6 & 5 \\
0 & 0 & 0 & 0 & 5 & 6 & 8 & 10 & 3
\end{array}
\right] .
\]
The degeneracy vector is $\mathbf{d=}(d_1,\ldots ,d_{14})\mathbf{=}(1,35,$ $%
28,112,$ $168,56,$ $112,24,$ $504,168,$ $168,336,$ $168,168).$ The entropy
exchange is
\begin{equation}
S(\rho _e)=-\sum_{i=1}^{14}d_i\xi _i\log _2(\xi _i).
\end{equation}
The eigenvalues of the output state $\rho ^{\prime }$ are
\begin{eqnarray}
\lambda _1 &=&\frac 18(\xi _1+35\xi _2+28\xi _3),  \nonumber \\
\lambda _2 &=&4(\xi _4+\xi _5),  \nonumber \\
\lambda _3 &=&\frac 12(\xi _4+8\xi _5+7\xi _6),  \nonumber \\
\lambda _4 &=&\frac 12(\xi _4+\xi _5+14\xi _7),  \nonumber \\
\lambda _5 &=&\frac 18[\xi _8+7(3\xi _9+\xi _{10}  \nonumber \\
&&+\xi _{11}+2\xi _{12}+\xi _{13}+\xi _{14})],
\end{eqnarray}
with degeneracy vector $\mathbf{d}_\rho ^{\prime }=8\mathbf{d}^{\prime },$
and $\mathbf{d}^{\prime }\mathbf{=}(d_1^{\prime },\ldots ,d_5^{\prime })%
\mathbf{=}(1,3,2,2,24).$ Thus the entropy of $\rho ^{\prime }$ is
\begin{equation}
S(\rho ^{\prime })=-8\sum_{i=1}^5d_i^{\prime }\lambda _i\log _2(\lambda _i).
\end{equation}
The average coherent information per channel use is
\begin{equation}
\overline{I_c}=-\sum_{i=1}^5d_i^{\prime }\lambda _i\log _2(\lambda _i)+\frac
18\sum_{i=1}^{14}d_i\xi _i\log _2(\xi _i).
\end{equation}
The coherent information is shown in Figure 1 and Figure 2.

\begin{figure}[tbp]
\includegraphics[width=3in]{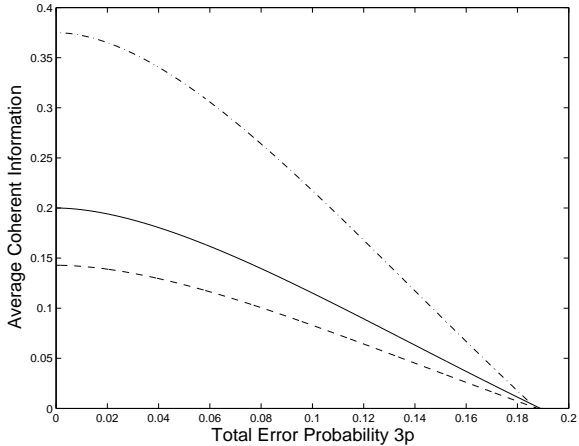}
\caption{Solid line for [[5,1,3]]; Dash line for [[7,1,3]]; Dot-dashed line
for [[8,3,3]]. }
\end{figure}

\begin{figure}[tbp]
\includegraphics[width=3in]{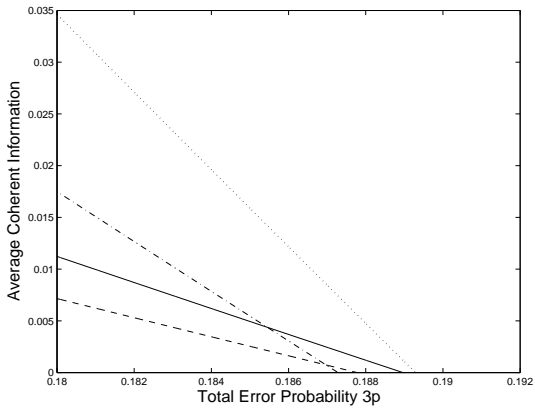}
\caption{Solid line for [[5,1,3]]; Dash line for [[7,1,3]]; Dot-dashed line
for [[8,3,3]]; Dot line for hashing bound.}
\end{figure}

\section{General Pauli channels}

For Pauli channel with $3$ error probabilities $p_x,$ $p_y,$ $p_z$, the
coherent information of coded input can be evaluated in the same way as
depolarizing channel. The only difference is that now we should count the
numbers of the each type of error separately in calculating $\sum_a\eta _a$.
Define the functions
\begin{eqnarray*}
g_1(t,z,x,y) &=&t^5+5t(x^2y^2+x^2z^2+z^2y^2), \\
g_2(t,z,x,y) &=&t^4x+2t^2x(z^2+y^2) \\
&&+2tzy(2x^2+z^2+y^2) \\
&&+x(x^2y^2+x^2z^2+z^2y^2).
\end{eqnarray*}
Then the eigenvalues of the joint output state $\rho _e$ are given in Table
1,

\begin{table}[tbp]
{\bfseries Table 1: The eigenvalues of $\rho _e$}\\[1ex]
\begin{tabular}{l|l|l|l}
\hline
$i$ & $\xi_{i1}$ & $\xi_{i2}$ & $d_{i} $ \\ \hline
1 & $g_1(f,p_z,p_x,p_y)$ & $g_1(p_y,p_x,p_z,f)$ & 1 \\
2 & $g_2(f,p_z,p_x,p_y)$ & $g_2(p_y,p_x,p_z,f)$ & 5 \\
3 & $g_2(f,p_z,p_y,p_x)$ & $g_2(p_y,p_x,f,p_z)$ & 5 \\
4 & $g_2(f,p_x,p_z,p_y)$ & $g_2(p_y,p_z,p_x,f)$ & 5 \\ \hline\hline
$i$ & $\xi_{i3}$ & $\xi_{i4}$ & $d_{i} $ \\ \hline
1 & $g_1(p_z,f,p_y,p_x)$ & $g_1(p_x,p_y,f,p_z)$ & 1 \\
2 & $g_2(p_z,f,p_y,p_x)$ & $g_2(p_x,p_y,f,p_z) $ & 5 \\
3 & $g_2(p_z,f,p_x,p_y)$ & $g_2(p_x,p_y,p_z,f) $ & 5 \\
4 & $g_2(p_z,p_y,f,p_x) $ & $g_2(p_x,f,p_y,p_z)$ & 5 \\ \hline
\end{tabular}
\\[0.5ex]
\newline
\end{table}
where the eigenvalues are expressed as $\xi _{ij}$ with degeneracy $d_i$.
The entropy exchange then is
\begin{equation}
S(\rho _e)=-\sum_{i=1}^4d_i\sum_{j=1}^4\xi _{ij}\log _2\xi _{ij}.
\end{equation}
The eigenvalues of $\rho ^{\prime }$ are $\lambda _i=\frac 12\sum_{j=1}^4\xi
_{ij}$ with degeneracy $2d_i$, thus we have
\begin{equation}
S(\rho ^{\prime })=-\sum_{i=1}^42d_i\lambda _i\log _2\lambda _i.
\end{equation}
The coherent information per channel use for Pauli channel with $[[5,1,3]]$
code as input is
\begin{eqnarray}
\overline{I_c} &=&-\frac 15\sum_{i=1}^4d_i[(\sum_{j=1}^4\xi _{ij})\log
_2(\frac 12\sum_{j=1}^4\xi _{ij})  \nonumber \\
&&-\sum_{j=1}^4\xi _{ij}\log _2\xi _{ij}].
\end{eqnarray}

\section{Dicussions and Conclusions}

There are bounds on the coding rate of QECC, the quantum Hamming
bound\cite {Ekert}, Knill-Laflamme (quantum Singleton) bound
\cite{Knill} , Gottesman bound and so on \cite{Gottesman}. The
first two give rather tight upper bounds on some of additive
quantum codes. The quantum Hamming bound (hashing bound) is a
strict upper bound for non-degenerate (pure) quantum code, as it
is seen from figure 2, where all three average coherent
information calculated are upper bounded by the hashing bound.
However, it has been known that quantum Hamming bound can be
violated by degenerate (impure) quantum codes \cite{DiVincenzo}.
They obtain the result by calculating the coherent information of
depolarizing channel with repetition quantum codes. We have
introduced a systematical way of calculating the coherent
information of Pauli channel with quantum code as input state. The
main finding of this paper is that the channel output density
matrix as well as the density matrix of the joint output of the
system and the auxiliary can be diagonalized for Pauli environment
with quantum code as input, the eigenvalue problem is reduced to
counting the weight of the error operators in the coset. We have
presented the input of $[[8,3,3]]$ code as an example of
calculating the coherent information of input state with multiple
logical qubits. It is anticipated that our method should promote
the way of violating quantum Hamming bound by calculating the
coherent information of depolarizing channel with quantum code of
encoding multiple logical qubit. Meanwhile, our method provide the
way of calculating the lower bound for the distillable
entanglement of quantum code state passing through Pauli channel,
according to hashing inequality\cite{Horodecki}.

Funding by the National Natural Science Foundation of China (Grant No.
60972071), Zhejiang Province Science and Technology Project (Grant No.
2009C31060) are gratefully acknowledged.

\end{document}